\documentclass[
	aps,
	prl,
	twocolumn, 		
	superscriptaddress, 	
	amsmath,
	showpacs,		
	showkeys,		        
	floatfix,
	a4paper,
	balancelastpage	
]{revtex4-1}

\usepackage[latin1]{inputenc}
\usepackage{graphicx}
\usepackage{dcolumn}
\usepackage{bm}
\usepackage{setspace}   

\usepackage{units}
\usepackage{color,soul}  

\newcommand {\Fig} {Fig.~}  
\newcommand{\onetwoten} {1\nobreakdash-2\nobreakdash-10}

\newcommand {\V}[1] {$\bm{#1}$} 
\newcommand {\etal} {\textit{et al.}}

\newcommand {\qaf} {\ensuremath{\bm{q}_{\mathrm{AF}}}}

\newcommand {\Tz} {\ensuremath{T_{0}}}

\newcommand {\CeRuAl} {CeRu$_{2}$Al$_{10}$}
\newcommand {\CeOsAl} {CeOs$_{2}$Al$_{10}$}
\newcommand {\CeFeAl} {CeFe$_{2}$Al$_{10}$}

\newcommand {\YbB} {YbB$_{12}$}

\begin{document}
\title{Dispersive magnetic resonance mode in the Kondo semiconductor  \CeFeAl\
}

\author{Jean-Michel Mignot}
\email[e-mail address: ]{jean-michel.mignot@cea.fr}
\affiliation{Laboratoire L\'{e}on Brillouin, CEA-CNRS, CEA/Saclay, 91191 Gif sur Yvette (France)}
\author{Pavel A. Alekseev}
\affiliation{National Research Centre `Kurchatov Institute', 123182 Moscow, Russia}
\affiliation{National Research Nuclear University ``MEPhI'', Kashirskoe sh. 31, 115409, Moscow, Russia}
\author{Julien Robert}
\author{Sylvain Petit}
\affiliation{Laboratoire L\'{e}on Brillouin, CEA-CNRS, CEA/Saclay, 91191 Gif sur Yvette (France)}
\author{Takashi Nishioka}
\author{Masahiro Matsumura}
\affiliation{Graduate School of Integrated Arts and Science, Kochi University, Kochi 780-8520 (Japan)}
\author{Riki Kobayashi}
\affiliation{Neutron Science Laboratory, Institute for Solid State Physics, University of Tokyo, Tokai, 319-1106, Japan}
\altaffiliation{Present Address: Quantum Condensed Matter Division, ORNL, Oak Ridge, Tennessee 37831, USA}
\author{Hiroshi Tanida}
\author{Hiroki Nohara}
\author{Masafumi Sera}
\affiliation{Department of Quantum Matter, ADSM, Hiroshima University, Higashi-Hiroshima, 739-8530 (Japan)}

\date{\today}

\begin{abstract}
The Ce$T_2$Al$_{10}$ family of orthorhombic compounds exhibits a very peculiar evolution from a Kondo-insulator ($T$: Fe) to an unconventional long-range magnetic order  ($T$: Ru, Os). Inelastic neutron scattering experiments performed on single-crystal \CeFeAl\ reveal that this material develops a spin-gap in its magnetic spectral response below $\sim 50$ K, with a magnetic excitation dispersing from $E = 10.2 \pm 0.5$ meV at the $Y$ zone-boundary point [$\bm{q} = (0, 1, 0)$] to $\approx 12$ meV at the top of the branch. The excitation shows a pronounced polarization of the magnetic  fluctuations along $a$, the easy anisotropy axis. Its behavior is contrasted with that of the (magnonlike) modes previously reported for \CeRuAl, which have transverse character and exist only in the antiferromagnetic state. The present observation is ascribed to a ``magnetic exciton'' mechanism invoked to explain a similar magnetic response previously discovered in \YbB.
\end{abstract}



\maketitle

Kondo insulators (KI) form a unique class of materials, in which semiconducting properties develop on cooling as a result of strong correlations between electrons from inner (usually 4$f$ or 5$f$) atomic shells. Hybridization of these originally localized orbitals with conduction-band states can produce the opening of a very narrow gap in the electronic density of states at the Fermi energy, which is responsible for a number of unconventional magnetic and transport properties \cite{Risebg'00}. Despite extensive investigations, the physics of these materials still conceals many riddles, as exemplified by the recent discovery of a so far unsuspected ``topological Kondo insulator'' phenomenon in the archetype KI compound SmB$_6$, with robust conducting states occurring at the surface of a truly insulating bulk \cite{* [{}] [{ and refs. therein.}] Reich'12}.

Another challenging issue in this field is the interplay between the nonmagnetic singlet ground state prevailing in most KI materials at low temperature, and the tendency to develop short-range, dynamical, antiferromagnetic (AF) correlations revealed by inelastic neutron scattering (INS) measurements. In \YbB, in particular, we have shown previously that the low-energy magnetic response in the KI state is dominated by a sharp, resolution-limited peak, located just below the edge of a spin gap, which disappears rapidly upon heating as the system enters the incoherent spin fluctuation regime \cite{Mignot'05, *Nemkovski'07}. This peak was suggested by Riseborough \cite{Risebg'01} to result from an exciton-type mechanism (a resonance mode in the spin response function), reflecting residual AF interactions between the renormalized 4$f$ quasiparticles. His interpretation is reminiscent of that proposed for the well-known ``resonance mode'' (RM) in high-$T_{c}$ superconductors \cite{*[{See e.g.: }] [{ and references therein.}] Eschrig'06}. A similar situation possibly occurs in SmB$_6$ as well \cite{Aleks'95}. On the other hand, evidence is still lacking for the existence of a RM-type excitation in the case of the Ce-based KIs, primarily because detailed studies of the \V{q} dependence of the magnetic response on single crystals are scarce \cite{Adroja'08}.

In this paper, we present new INS results showing that \CeFeAl\ is likely to be the first example of a Ce system with a RM excitation in the KI state. This compound belongs to the ``\onetwoten'' series ($RT_2$Al$_{10}$ with $R$ a lanthanide element and $T$ a transition metal element such as Fe, Ru or Os). Unlike its Ru and Os counterparts, which have been extensively investigated for their unconventional combination of KI behavior and long-range magnetic order \cite{Muro'09, *Nishioka'09, *Strydom'09}, \CeFeAl\ appears to be a rather classical paramagnetic KI, as far as magnetic and transport properties are concerned. The electrical resistivity measured along the $a$ axis exhibits an overall negative temperature coefficient (except for a broad maximum occurring near $T^{a}_{max} = 70$ K) and saturates below $T = 1$ K at a fairly high value of $\approx 1300\ \mu \Omega \mathrm{cm}$. In contrast to cubic boride KIs, orthorhombic \CeFeAl\ is strongly anisotropic, with a paramagnetic susceptibility ratio between $\chi_a$ (easy axis) and $\chi_b$ (hard axis) reaching 5.2 at $T^{a}_{max}$. In a recent work, Adroja and coworkers \cite{Adroja'13} studied the spin dynamics in Ce(Ru$_{1-x}$Fe$_{x}$)$_2$Al$_{10}$ solid solutions, using time-of-flight (TOF) INS experiments on powder samples. For $x = 1$ (pure \CeFeAl), the magnetic response at $T = 7$ K shows no intensity below 5 meV, but a pronounced low-energy peak is observed at about 13 meV, which is taken to represent the energy of the spin gap. However, powder experiments cannot reveal where the signal occurs in \V{Q} space nor whether or not it exhibits sizable dispersion. These questions are central to establishing the exact nature of the excitation, and we have thus performed detailed measurements of the magnetic spectral response in \CeFeAl\ using a single-crystal sample. The results are reported in the following.
 
Fourteen single crystals of \CeFeAl\ (base-centered orthorhombic, $Cmcm$ space group, No. 63, $a = 9.0159$ \AA, $b = 10.2419$ \AA, $c = 9.0882$ \AA\ \cite{Sera'13}) with dimensions comprised between 1 and 7 mm, for a total mass of about 700 mg, were grown by an Al-flux method. they were then co-aligned on an Al sample holder, using Cytop CTL-809M fluoropolymer glue with low hydrogen content. A total mosaic spread of about 3 degrees was estimated from the neutron rocking curves around the $c^*$ axis, which was sufficient for the present experiment. Excitation spectra were measured in the $(a^*, b^*)$ scattering plane on the 2T triple-axis spectrometer (TAS) at LLB-Orph\'ee (Saclay). Spectra were recorded at fixed final energy, $E_f=14.7$ meV, using a pyrolytic graphite (PG) 002 monochromator and analyzer, with a PG filter placed on the scattered beam. The instrumental resolution was $\approx 1.1$ meV at the elastic position and $\approx 1.7$ meV for $E = 10$ meV.

\begin{figure} [!b] 	
\includegraphics [width=0.85\columnwidth] {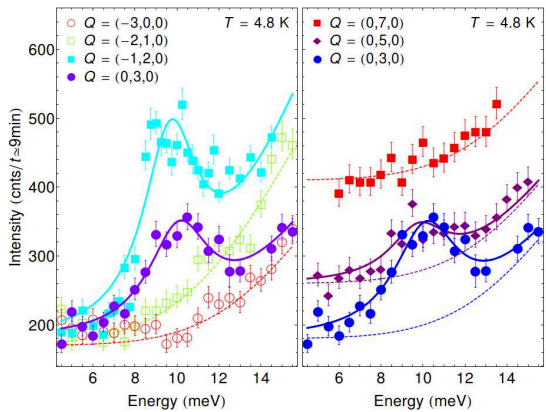} 
\caption{\label{fscans_Y} (Color online) Energy scans at constant neutron final energy $E_f = 14.7$ meV, measured at $T = 4.8$ K for momentum transfers corresponding to the same reduced \V{q} vector $(0, 1, 0)$ (zone boundary Y point). The left frame demonstrates the dependence of the inelastic peak intensity on the direction of \V{Q} with respect to the $a$ axis (dipole selection rule). The right frame shows the rapid suppression of the peak with increasing momentum transfer. Dashed lines: estimated background (see text). Solid lines: fits assuming Lorentzian lineshapes.}  
\end{figure}

Constant-\V{Q}\ scans recorded at the base temperature, $T_{\mathrm{min}} \approx 4.8$ K, are presented in \Fig \ref{fscans_Y}. For \V{Q} = (0, 3, 0) and (-1, 2, 0), one sees a clear peak, centered at $E \approx 9.5$ meV, superimposed on a rather large sloping background. A survey of its intensity in \V{Q} space indicates that this signal is peaked near the reduced \V{q} vector (0, 1, 0)  ($Y$ point in Brillouin zone), which corresponds to the wavevector of the AF order previously reported for \CeRuAl\ or \CeOsAl\ \cite{Khalyavin'10,Robert'10,*Mignot'11}. Measurements at larger scattering vectors, for \V{Q} = (0, 5, 0) and (0, 7, 0) (right frame), show that the peak intensity is rapidly suppressed with increasing momentum transfer, as expected for magnetic scattering. Meanwhile, the background increases significantly, suggesting that it originates, at least partly, from phonon nuclear scattering. This general pattern is qualitatively consistent with the powder results of Ref.~\cite{Adroja'13}, which showed an excitation at about 13 meV, partly overlapping the low-energy tail of a broad peak centered near 50 meV. 

\begin{figure} [!b] 	
\includegraphics [width=0.60\columnwidth] {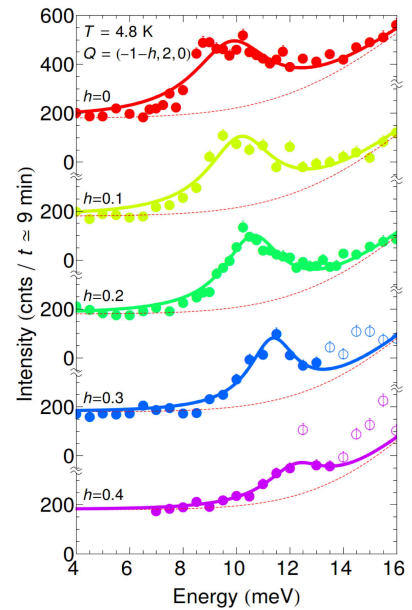} 
\caption{\label{fscans_disp} (Color online) Energy scans at constant neutron final energy $E_f = 14.7$ meV, measured at $T = 4.8$ K for momentum transfers $\bm{Q} = (-1-h, 2, 0)$, showing the dispersion of the magnetic excitation and the \V{Q} dependence of its intensity along the $(h, 0, 0)$ direction. Dashed lines: estimated background (see text). Solid lines: fits assuming Lorentzian lineshapes; open symbols denote data points thought to contain a contamination and thus excluded from the fit.}  
\end{figure}

The difference between the energy of the peak in the present experiment and in the powder-averaged data suggests that the excitation may show some dispersion. To check this assumption, scans have been performed along several high-symmetry directions in the $(a^{\ast}, b^{\ast})$ scattering plane. The spectra exhibit a positive dispersion starting from the $Y$ points \V{Q} = (0, 3, 0) and (-1, 2, 0) (see, e.g. \Fig \ref{fscans_disp}). Meanwhile, the intensity decreases rapidly and the peak becomes barely detectable when \V{q} departs from the $Y$ point by more than 0.4 r.l.u. in any direction. To derive the \V{Q} dependence of the excitation energy, linewidth, and integrated intensity, the data were fitted assuming a Lorentzian line shape (detailed balance factor corrections are insignificant at this temperature for the energy range of interest) \footnote{Results obtained when using Gaussian fits do not differ qualitatively from those presented here}. The main source of uncertainty is the determination of the background, especially at the high-energy end of the scan where some contamination seems to exist \footnote{High-orders of the Bragg reflections on the monochromator and analyzer are known to cause extra scattering for $E = 18.35$ meV ($2k_i = 3k_f$). That signal may appear shifted in energy, in the case of coherent scattering, because of resolution effects}. This background is comparably steep for all \V{Q} vectors investigated, but not strictly identical. It was therefore estimated from a comparison of low- and high-temperature spectra for the same \V{Q} vector, assuming the latter to consist of a single broad quasielastic (QE) contribution varying slowly in \V{Q} space (see below). 

\begin{figure} [!t] 	
\includegraphics [width=0.65\columnwidth] {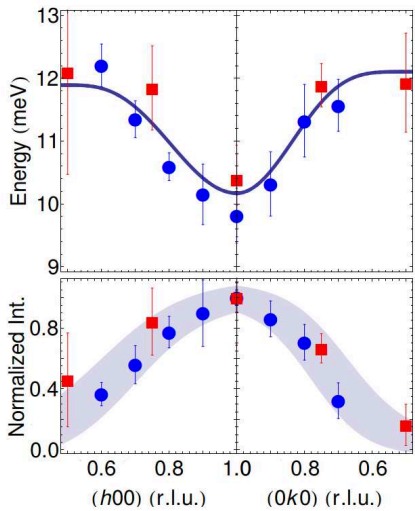} 
\caption{\label{fdisp} (Color online) Experimental dispersion curve and \V{Q} dependence of the intensity of the magnetic excitation at $T \approx 4$ K. Red squares [for $\bm{Q} = (h, 3, 0)$ and $(0, 3 - k, 0)$] and blue circles [for $\bm{Q} = (-1 - h, 2, 0)$ and $(0, 2 + k, 0)$] correspond to spectra measured along equivalent, yet distinct, $(h, 0, 0)$ (left frame) and $(0, k , 0)$ (right frame) directions in reciprocal space.}  
\end{figure}

\begin{figure} [!t] 	
\includegraphics [width=0.95\columnwidth] {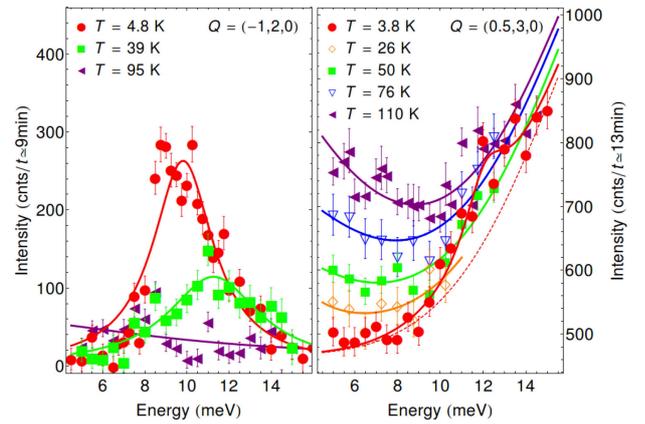} 
\caption{\label{ftdep} (Color online) Temperature evolution of the magnetic spectral response measured at constant neutron final energy Ef = 14.7 meV for $\bm{Q} = (-1, 2, 0)$ (left) and $(0.5, 3, 0)$, showing the rapid suppression of the inelastic peak and the gradual appearance of the quasielastic signal as $T$ increases. In order to improve statistics, the scans denoted 50 K, 95 K, and 110 K were obtained by combining data measured at 45 K and 55 K, 80 K and 110 K,  100 K and 120 K, respectively.}  
\end{figure}

The results are plotted in \Fig \ref{fdisp}, where the error bars account for the above-mentioned uncertainties. One sees that the dispersion, although less pronounced than in the AF phase of \CeRuAl\ \cite{Robert'12}, is still significant. Starting from a minimum slightly in excess of 10 meV at the $Y$ point, it reaches about 12--13 meV near the top of the branch. This means that the excitation energy of 13 meV obtained in Ref.~\cite{Adroja'13} from the energy of the peak in the TOF spectra was overestimated, likely because of the minor weight, in the \V{Q}-space average, of the region close to the $Y$ point where the minimum is located. A similar disagreement between the TOF and TAS results was already pointed out in Ref.~\cite{Robert'12} for \CeRuAl. Within experimental accuracy, the present dispersion appears to be similar along the three symmetry directions investigated [$(h, 0, 0)$ and $(0, k, 0)$ in \Fig \ref{fdisp}, $(h, -h, 0)$ not shown]. The width of the peak (full width at half maximum of about $4.0 \pm 0.7$ meV), does not vary significantly between the different \V{Q} vectors where it was measured, and remains much larger than that expected from the experimental resolution (1.7 meV at the peak energy). The asymmetric shape of the (-1, 2, 0) peak, after subtracting the background (\Fig \ref{ftdep}, left frame), could indicate a doublet of excitations but, in view of the insufficient statistics and the possibility of contaminations, this possibility could not be confirmed.

The temperature dependence of the excitation, measured for \V{Q} = (-1, 2, 0), is plotted in \Fig \ref{ftdep} (left). The peak intensity decreases rapidly on heating to 40 K, and vanishes at 80 and 110 K. Meanwhile, intensity appears gradually at low energy (right frame), reflecting the growth of QE fluctuations. This signal can be fitted to a Lorentzian peak, with a half width at half maximum of about 5 meV showing little temperature dependence between 45 K and 120 K. \footnote{In Ref.~\cite{Adroja'13}, a much larger QE linewidth was obtained from TOF spectra at $T = 300$ K. However, the data covered a  wide energy range up to 80 meV. In higher-resolution spectra (measured with $E_i = 20$ meV), extra intensity is seen to exist at low energy, which could correspond to the signal observed in the present measurements.}

It is interesting to note that, at $T = 40$ K, the inelastic peak is strongly suppressed without shifting to lower energy (the data even suggest a slight increase in the excitation energy but this observation needs to be confirmed). The appearance of a signal below 8 meV for $T > 50$ K is therefore ascribed to a filling, rather than a closing, of the spin gap. This behavior is at variance with that reported for \CeRuAl\ \cite{Robert'10, Khalyavin'10, Mignot'14p}, in which the energy of the magnonlike excitation at the spin-gap edge decreases as $T$ increases toward the N\'eel temperature \Tz\ in the AFM phase. On the other hand, it is reminiscent of the temperature suppression of the magnetic exciton mode in \YbB\ \cite{Mignot'05, Nemkovski'13}. 

Information on the polarization of the AF fluctuations can be obtained from the \V{Q} dependence of the dynamic response, using the dipole selection rule for magnetic neutron scattering, which stipulates that only magnetic components perpendicular to the scattering vector contribute. Here the comparison of the spectra for \V{Q} = (-1, 2, 0)  and (0, 3, 0) on the one hand, \V{Q} = (-2, 1, 0)  and (3, 0, 0) on the other hand, clearly shows that the peak is suppressed (or strongly reduced)  when the scattering vector is oriented along (or close to) the $a^{\ast}$ direction (\Fig \ref{fscans_Y}, left frame). This implies that the signal mainly arises from correlations $\langle m_i^a m_j^a \rangle$ between magnetic components parallel to $a$, which is also the easy axis observed in static magnetization measurements \cite{Nishioka'09}.

In the QE regime, the signal at low energy is observed with sizable intensity at various momentum transfers, irrespective of their orientation in \V{Q} space (at least within the experimental (001) scattering plane). This is taken to indicate that the predominance of magnetic correlations along $a$ observed at $T = 5$ K is a genuine feature of the KI state, which does not extend into the incoherent relaxational regime. Additional measurements as a function of temperature at different \V{Q} vectors should be carried out to trace this anisotropy in more detail.

The present study shows that the low-temperature spin dynamics in \CeFeAl\ is quite unconventional for a Ce-based KI compound and, in particular, differs markedly from that observed previously in \CeRuAl. In the latter compound, the spin-gap response, associated with well-defined magnetic excitation branches, was essentially restricted to the AF phase, and could be explained using a RPA model for anisotropic AF magnons \cite{Robert'12}. In the present case, on the other hand, the dispersive mode exists \textit{in the absence of long range magnetic order}, and its appearance rather seems to correlate with the low-temperature upturn in the resistivity $\rho_{\parallel}$$_a$ below $\sim\ 50$ K, which signals the entry into the KI regime. This point is important because, despite undisputed evidence for a long-range AF ordered phase in \CeRuAl\ (as well as \CeOsAl) below \Tz, there is ongoing controversy as to whether the spin gap in those systems should be interpreted as the anisotropy gap of magnonlike excitations, or in terms of a singlet-triplet level scheme due to a spin-dimer or charge-density-wave transition occurring somewhat above \Tz\ \cite{Adroja'13}. 

Since the AF correlations giving rise to the inelastic peak in \CeFeAl\ are centered at the same $k$ vector where the AF structure develops in \CeRuAl, they can be regarded as a precursor effect of the long-range order, which could develop, e.g., under application of a negative pressure. However, unlike the QE signal existing in \CeRuAl\ over a limited temperature interval above \Tz, which exhibits both a \V{Q}-dependent intensity, peaked around \qaf, and correlations polarized along $a$ \cite{Mignot'14p}, the QE response in \CeFeAl\ above 50 K is widely spread in \V{Q} space and shows no evidence of mode polarization. With lowering the temperature, this local paramagnetic relaxation regime evolves toward the appearance of the spin gap and the inelastic mode, contrasting with \CeRuAl\ in which AF critical fluctuations develop on approaching \Tz.

Magnetic anisotropy is known to play a major role in \onetwoten\ compounds, because of their particular crystal structure. The model used to describe the  excitations in \CeRuAl, and even to account for the orientation of the ordered AF moments below \Tz, actually requires strongly anisotropic exchange constants. However, the mode polarization found here for \CeFeAl\ is likely dominated by the single-ion (crystal-field) anisotropy and gives no indication as to the exchange interactions.  

From the above results, the spin dynamics in \CeFeAl\ is seen to bear close similarities with the ``exciton'' response previously reported for \YbB. In both systems a specific magnetic excitation develops in the KI regime, whose position is close to the spin-gap edge. The mode exhibits a positive dispersion from a zone-boundary \V{q} vector, with intensity decreasing rapidly as the energy comes closer to the continuum. With increasing temperature, the peak is rapidly suppressed but does not appear to move significantly to lower energies, as was already noted for \YbB\ \cite{Nemkovski'13}. According to Riseborough \cite{Risebg'01}, the exciton peak represents a ``resonance mode'' due to residual AF interactions between quasiparticles in the periodic Anderson model. It requires both a sizable exchange interaction $J(\qaf)$ and a large value of the magnetic susceptibility $\chi_0(\qaf,E)$. In his calculation the latter quantity, associated with particle-hole excitations, is specifically enhanced for a \V{q} vector corresponding to that of the (indirect) hybridization gap, because Kramers-Kronig relation produces a peak in the real part of $\chi_0(\qaf,E)$ at the spin-gap edge, where the imaginary part rises steeply. An important consequence of this picture is that the peak energy can no longer be identified with 
the spin-gap value, as is often done in experimental studies of KIs \cite{Adroja'08}, since the RM is essentially an in-gap excitation. In practice, however, the edge of the electron-hole continuum may be difficult to ascertain from the measured spectra.

Unlike \YbB\ where the exciton peak was found to be resolution limited \cite{Mignot'05, *Nemkovski'07},  \CeFeAl\ exhibits sizable broadening of the inelastic peak, which is too large to be accounted for by experimental effects such as, e.g. mosaic spread in the crystal. This suggests that the damping is intrinsic and reflects a finite lifetime of the excitation. A possible origin for this effect is the residual density of state at the Fermi energy at 10 K inferred by Kimura \etal\ \cite{Kimura:'11.Fe} from the Drude weight remaining below 10 meV in their optical conductivity experiments.

In CeNiSn, one of the few KI compounds for which detailed single-crystal INS data exist, two magnetic excitations have been observed in the low-temperature regime for \V{Q} = $(Q_a, \nicefrac{1}{2} + n, Q_c)$ ($E = 4$ meV) and for \V{Q} = (0, 1, 0) and (0, 0, 1) ($E = 2$ meV), with $n$ integer and $Q{_a}$ and $Q{_c}$ arbitrary \cite{*[{}] [{ and refs. therein.}] Sato'95}. That compound is orthorhombic like  \CeFeAl\ but has a smaller gap ($\Delta \approx 6$ K) and a lower coherence temperature ($T_{\mathrm{coh}} \approx 12$--20 K). Cold-neutron TAS experiments \cite{Raymond'98}\ have shown that the magnetic spectral response has a pseudo gap, rather than a gap for \V{Q} = (0, 1, 0), since a sizable QE signal remains down to $T = 1.5$ K. This signal is \V{Q} independent, and therefore similar to that observed in \CeFeAl\ above 50 K. The  excitation at 2 meV has 3D character and a pronounced polarization along the easy $a$ axis, like in the present case, and disappears upon heating to 20 K. On the other had, no evidence was reported for a dispersion of that mode, which appears to be strongly peaked near the AF \V{q} vector. In Riseborough's model, some intermediate valence character of the material is considered favorable for stabilizing the in-gap mode, whereas CeNiSn is better categorized as a Kondo compound. The question whether the excitation can be ascribed to a spin-exciton mechanism thus remains open.

In view of the nonmagnetic singlet character of the \CeFeAl\ ground state, an analogy could also be searched with magnetic excitations (crystal-field excitons) occurring in Pr compounds with a single-ion singlet ground state and exchange-induced magnetic order. Contrasted behaviors were predicted---and experimentally observed, depending on the degeneracy of the excited state. In the \onetwoten\ compounds showing long range magnetic order, an electronic transition (dimerization, charge or spin density wave) has been suspected to occur above the Néel temperature, which could lead to a singlet-triplet level scheme. Such a situation was studied, for instance, in Pr$_3$Tl \cite{Birgeneau'71}, with the unexpected result that, unlike in the single-singlet case, the mode softening at the (ferro-) magnetic transition remains incomplete \cite{Smith'72}. The relevance of this picture to the spin-gap formation in \onetwoten\ KI compounds remains to be checked.

In summary, the present experiments reveal an unexpected \V{Q} dependence in the spin dynamics of \CeFeAl\ in the KI state. The inelastic peak first observed in TOF powder experiments is found to exhibit sizable dispersion but relatively weak anisotropy with respect to the direction in the $(a^{\ast}, b^{\ast})$ scattering plane. This excitation presents strong similarities with the ``resonance mode'' previously observed in the Kondo insulator \YbB, and is therefore suggested to arise from the same type of spin-exciton mechanism, originally proposed by Riseborough. However, in contrast to the latter cubic system, the AF fluctuations involved in this process exhibit a pronounced polarization along the $a$ axis. Application of the spin-exciton model to \CeFeAl\ will thus require proper treatment of the lower symmetry in \onetwoten\ compounds. 

\begin{acknowledgments}
We thank B. Annigh\"ofer for designing the sample holder and D. Adroja for providing Cytop glue. PAA's participation in this work was partly funded by the RFBR grant 14-02 00272-a, and by the French Ministry of Foreign Affairs (\textit{Séjours scientifiques de haut niveau})
\end{acknowledgments}


%

\end{document}